\documentclass[twocolumn,pre,showpacs,showkeys]{revtex4}
\usepackage{amsmath,amssymb}
\usepackage{graphicx}

\begin{document}

\title{Crystal vibration limit of nanoscale dynamic etalon. Rough estimate}
\author{Ihor Lubashevsky}
\affiliation{General Physics Institute, Russian Academy of
 Sciences, Vavilov Str. 38, Moscow, 119991 Russia}
\author{Peter Luskinovich }
\affiliation{General Physics Institute, Russian Academy of
 Sciences, Vavilov Str. 38, Moscow, 119991 Russia}

\date{\today }

\begin{abstract}

The paper analyzes the limits of nanoscale dynamic etalon caused by thermal
vibration of crystalline structure. It is shown that exactly sufficiently long
one-dimensional defect of crystal lattice should confine the standard accuracy
within angstroms.

\end{abstract}
\maketitle


One of the novel techniques of measuring objects of nanometer and subnanometer
scales is related to creating dynamic etalons based on the precise measurements
the tunnel microscope probe movements \cite{patent}. A linear dimension is
transmitted from the etalon to a measuring system by means of a controlled
movement of the etalon surface at a required distance and measurement of the
above movement. In this case the real accuracy attained by the measuring system
is limited by unpredictable vibrations of the crystal surface. The present
paper is devoted to analysis of the fundamental limits of such system caused by
thermal vibrations of crystal lattice.

The thermal deviation of crystal atoms from the equilibrium states is analyzed
within the adiabatic approximation assuming the electron polarization to follow
the relatively slow motion of atoms without delay. For molecular or covalent
crystals like diamond, Ge, Si, GaAs the electron polarization effect is
ignorable, whereas for ionic crystals (NaCl, CsCl, ZnS, etc.) this assumption
is well justified for all the wave vectors of phonons except for a sufficiently
small wave vectors $k_p\sim \Omega \sqrt{\epsilon}/c$ (see, e.g.
Ref.~\cite{Davydov,Zaiman}). Here $\Omega$ is a characteristic frequency of
optical phonons, $\epsilon$ is a crystal permittivity, and $c$ is the light
speed. In particular, for ZnS, where $\Omega \sim 6\times 10^{13}$~s$^{-1}$ and
$\epsilon\sim 7$, we get $k_p\sim 5\times 10^3$~cm$^{-1}$. By contrast,
according to the results to be obtained, actually phonons with wave vectors
about $k\lesssim 1/a\sim 2\times 10^{7}$~cm$^{-1}$ ($a$ is a characteristic
lattice spacing) mainly contribute to the characteristic values of atom thermal
fluctuations.

To estimate possible effects caused by the crystal surface states and its
defects we analyze 1D-, 2D- and 3D-models. It should be noted that in the cases
of 1D- and 2D-models the corresponding lattice is assumed to be embedded in the
3D-space, so its atoms actually perform 3D-motion.

Let the lattice unit cell contain $\sigma$ atoms and the value
$\mathbf{r}_{\mathbf{n}\alpha}$ specify the displacement of atom $\alpha$
($\alpha = 1,2,\ldots,\sigma$) in unit cell $\mathbf{n}=\{n_1,n_2,n_3\}$
($n_i\in \mathbb{Z}$) from its equilibrium position $\mathbf{R}_{\mathbf n} =
n_1\mathbf{a}_1 + n_2\mathbf{a}_2 + n_3\mathbf{a}_3$, where $\{\mathbf{a}_i\}$
are the crystalline spacing vectors (for the 2D-lattice $n_3 = 0$ and for the
atom chain $n_2 = 0$ also). Then the corresponding operator
$\hat{r}_{\mathbf{n}\alpha}$ and the conjugate creation--annihilation operators
$b^+_{\mathbf{k}s}$, $b^{\vphantom{+}}_{\mathbf{k}s}$  of phonons are related
by the expression (see, e.g., Ref.~\cite{Davydov})
\begin{equation}\label{eq:1}
    \hat{r}_{\mathbf{n}\alpha} = \sqrt{\frac{\hbar}{2Nm_\alpha}}\sum_{\mathbf{k},s}
    \frac{\mathbf{u}_{\alpha s}(\mathbf{k})}{\sqrt{\Omega_s(\mathbf{k})}}
    \left(b^{\vphantom{+}}_{\mathbf{k}s} + b^+_{-\mathbf{k}s} \right)
    e^{i\mathbf{kR_n}}\,.
\end{equation}
Here $N$ is the total number of unit cells, the sum runs over all the possible
wave vectors $\{\mathbf k \neq 0\}$ and all the modes, $s =
1,2,\ldots,3\sigma$, describing different vibrations of atoms within one unit
cell, $m_\alpha$ is the mass of atom $\alpha$, $\Omega_s(\mathbf{k})$ is the
frequency of the vibration type labeled by the pair $\{\mathbf{k},s\}$. The
eigenvectors $\mathbf{u}_{\alpha s}(\mathbf{k}) =\{u^{(l)}_{\alpha
s}(\mathbf{k})\}$ $(l = 1,2,3)$ of the corresponding vibration types meet the
equalities
\begin{equation}\label{eq:2}
    \sum_{\alpha = 1}^{\sigma}
    \mathbf{u}_{\alpha s}(\mathbf{k}) \cdot \mathbf{u}_{\alpha s'}(\mathbf{k})
    = \delta_{s,s'}\,,
\end{equation}
where $\delta_{s,s'}$ is the Kronecker symbol. Besides, the phonon frequency
meets the equality $\Omega_s(-\mathbf{k}) = \Omega_s(\mathbf{k})$, as well as
the eigenvectors do, $\mathbf{u}_{\alpha s}(\mathbf{-k})=\mathbf{u}_{\alpha
s}(\mathbf{k})$.

In order to estimate the characteristic amplitude $\delta h$ of atom thermal
fluctuations near the equilibrium states let us calculate the value
\begin{equation}\label{eq:3}
    \Big\langle\left(\delta h\right)^2\Big\rangle\sim
    \frac1{3\sigma}\,\Big\langle
    \sum_\alpha\left(\hat{r}_{\mathbf{n}\alpha}\right)^2
    \Big\rangle
\end{equation}
averaged over all the possible states of the equilibrium phonon ensemble with
the temperature $T$ and Hamiltonian
\begin{equation}\label{eq:4}
    \widehat{H} = \sum_{\mathbf{k},s}\Omega_s(\mathbf{k})
    \big[b^{+}_{\mathbf{k}s} b^{\vphantom{+}}_{\mathbf{k}s}+\tfrac12\big]\,.
\end{equation}
In this way we get
\begin{equation}
    \label{eq:5}
    \Big\langle\left(\delta h\right)^2\Big\rangle  \sim
    \frac{\hbar}{3N\sigma}\sum_{\mathbf{k},s}
    \frac{[n_s(\mathbf{k})+\frac12]}{M_s(\mathbf{k})\Omega_s(\mathbf{k})}\,,
\end{equation}
where the occupation number of phonons in the state $\big|\mathbf{k},s\big>$
under the thermodynamic equilibrium with temperature $T$ (measured in energy
units) is
\begin{equation}\label{eq:6}
    n_s(\mathbf{k})
    = \Big[ \exp\Big(\frac{\hbar\Omega_s(\mathbf{k})}{T}\Big)-1\Big]^{-1}
\end{equation}
and $M_s(\mathbf{k})$ is some effective mean mass of atoms in the unit cell
depending on the vibration mode that obeys the equality
\begin{equation}\label{eq:7}
    \frac1{M_s(\mathbf{k})} = \sum_\alpha\frac1{m_\alpha}
    \big[\mathbf{u}_{\alpha s}(\mathbf{k})\big]^2\,.
\end{equation}
Due to equality~\eqref{eq:2} the value $M_s(\mathbf{k})$ belongs to the
interval $M_s(\mathbf{k})\in [m_{\min}, m_{\max}]$, the boundary of the given
interval are specified by the minimal and maximal masses of the unit cell
atoms.

In estimating the value of $\langle(\delta h)^2\rangle$ let us ignore the
dependence of effective mass $M_s(\mathbf{k})$ on the wave vector $\mathbf{k}$,
i.e. set it equal to some constant, $M_s(\mathbf{k}) = M_{\text{eff}}$. Within
the adopted approximation sum~\eqref{eq:5} is reduced to the integral over the
possible values of the phonon energy $E=\hbar\Omega$:
\begin{equation}\label{eq:8}
    \sum_{\mathbf{k},s}(\ldots)\Rightarrow
    3N\sigma\int^{E_{\max}}_0 dE\rho(E)(\ldots)\,.
\end{equation}
Here the density of the phonon states $\rho(E)$ describes their distribution
over the interval from zero to some maximal value $E_{\max}$ and is to be
normalized to unity,
\begin{equation}\label{eq:9}
    \int^{E_{\max}}_0 dE\rho(E) = 1\,,
\end{equation}
because the total number of states $\big|\mathbf{k},s\big>$ must be equal to
$3N\sigma$. Thereby estimate~\eqref{eq:5} can be represented as
\begin{multline}\label{eq:10}
    \Big\langle\frac{\left(\delta h\right)^2}{a^2}\Big\rangle  \sim
    \frac{\hbar^2}{M_{\text{eff}}a^2}
    \int^{E_{\max}}_0 dE\rho(E)\\
    \times\frac1{E}
    \left[\frac1{\exp\big(\frac{E}{T}\big)-1}+\frac12\right]\,,
\end{multline}
where, as before, $a$ denotes the characteristic value of the crystal lattice
spacing.

The upper boundary of the phonon energies $E_{\text{max}}$ can be estimated
using the characteristic frequency of the optical phonons $\Omega_{\text{opt}}$
or what is the same the characteristic frequency of individual atom vibrations
in crystal lattices, $E_{\text{max}}\sim \hbar\Omega_{\text{opt}}$ (see, e.g.,
\cite{Davydov}). For example, dealing with ZnS we have $\Omega_{\text{opt}}\sim
6\times 10^{13}$~s$^{-1}$, which corresponds to temperatures about $T_D\sim
\hbar\Omega_\text{opt}/k_{B}\sim 460$~K.

To be specific in what follows we confine our consideration to the case of not
too high temperatures, $T\lesssim T_D$. Besides, setting $M_\text{eff}\sim 40$
atomic units (for ZnS atomic masses of the compounds are $M_\text{Zn}\approx
69$ and $M_\text{S}\approx 32$) and $a\sim 3 {\AA}$ the following estimates of
the cofactor in formula~\eqref{eq:10}
\begin{equation}\label{eq:11}
    \Re:=
    \frac{\hbar^2}{M_{\text{eff}}a^2}\cdot\frac1{\hbar \Omega_\text{opt}}\sim
    3\times 10^{-4}
\end{equation}
is obtained. Therefore it is possible to ignore the effect of quantum
fluctuations (the summand 1/2 in expression~\eqref{eq:10}), so
\begin{equation}\label{eq:12}
    \Big\langle\frac{\left(\delta h\right)^2}{a^2}\Big\rangle  \sim
    \frac{\hbar^2}{M_{\text{eff}}a^2}
    \int\limits^{E_{\max}}_0 dE
    \frac{\rho(E)}{E\big[\exp\big(\frac{E}{T}\big)-1\big]}\,.
\end{equation}

If the phonon spectrum would be characterized by a single frequency $\Omega$,
i.e. $\rho(E) = \delta(E - \hbar \Omega)$, then the desired estimate has taken
the form
\begin{align}
    \nonumber
    \Big\langle\frac{\left(\delta h\right)^2}{a^2}\Big\rangle & \sim
    \frac{\hbar}{M_{\text{eff}}a^2\Omega}
    \Big[\exp\Big(\frac{\hbar\Omega}{T}\Big)-1\Big]^{-1}\\
    \label{eq:13}
    & \sim
    \frac{T}{M_{\text{eff}}a^2\Omega^2}\qquad
    \text{for}\quad T \gtrsim \hbar\Omega\,.
\end{align}
This expression actually corresponds to the model of a disconnected oscillator
ensemble and the lower line of expression~\eqref{eq:13} matches the estimate of
the amplitude of classical oscillator affected by the white noise with
intensity $T$. For $T\sim \hbar\Omega$ and $\Omega\sim \Omega_\text{opt}$ the
value of the given estimate is evaluated by expression~\eqref{eq:11}.  It is
actually an estimate of the thermal vibration amplitude of a point defect in
the bulk or surface of the crystal. So their effects on the accuracy of the
nanoscale dynamical standard may be ignored.

In order to get a more realistic estimate we will make us of the Debay model
approximating the real phonon spectrum with solely acoustic phonons where
$\Omega_s(\mathbf{k})\propto k$. It is equivalent to approximating the phonon
density $\rho$ as follows
\begin{equation}\label{eq:14}
    \rho(E) = D\frac{E^{D-1}}{E_\text{max}^{D}}
\end{equation}
where $D$ is the dimension of the crystal lattice under consideration,
$D=1,2,3$. Then, taking into account the adopted assumption about the value of
temperature, $T\lesssim T_D$ (recall that $k_B T_D := E_\text{max}: = \hbar
\Omega_D$), we can set the upper boundary of integral~\eqref{eq:12} to infinity
and, thus, rewrite expression~\eqref{eq:12} as
\begin{equation}\label{eq:15I}
    \Big\langle\frac{\left(\delta h\right)^2}{a^2}\Big\rangle  \sim
    D \Re
    \left(\frac{T}{T_D}\right)^{D-1}
    \int\limits^{\infty}_0 dx
    \frac{x^{D-1}}{x\big[\exp(x)-1\big]}\,.
\end{equation}
For the 3-dimensional lattice crystal integral~\eqref{eq:15I} is reduced to the
following
\begin{equation}
    \label{eq:15D3}
   \Big\langle\frac{\left(\delta h\right)^2}{a^2}\Big\rangle  \sim
    \frac{\pi^2\Re}{2}
    \Big(\frac{T}{T_D}\Big)^{2}\,,
\end{equation}
i.e. again thermal vibrations of solid bulk has a minor effect on the accuracy
of dynamical nanostandart.

For the 2-dimensional (plane) lattice integral~\eqref{eq:15I} has a formal
logarithmic singularity that has to be cut off at small energies related to
phonons with very long wavelengths bounded by the crystal size or surface
structures. It leads to appearing some cofactor $L_n$ in the expression
\begin{equation}
    \label{eq:15D2}
   \Big\langle\frac{\left(\delta h\right)^2}{a^2}\Big\rangle  \sim
    \Re L_n
    \Big(\frac{T}{T_D}\Big)\,,
\end{equation}
but its value cannot be too large, typically such effects are reduced to
$L_n\sim 10$. So, again, possible phonon modes localized at crystal surface
which have be simulated with the 2-dimensional lattice, has an insignificant
effect of the standard accuracy.

For the 1-dimensional chain the situation changes dramatically.
Integral~\eqref{eq:15I} has a singularity of the type $1/x$ and its long-wave
vibrations give rise to significant deviation of atoms from the equilibrium
positions, i.e.
\begin{equation}
    \label{eq:15D1}
   \Big\langle\frac{\left(\delta h\right)^2}{a^2}\Big\rangle  \sim
    \Re    \Big(\frac{T\ell}{\hbar c_s}\Big)\,,
\end{equation}
because the lower boundary of the phonon energy can be estimated as $\hbar
c_s/ell$ where $c_s\sim 10^5$~cm/s is sound velocity in solids and $\ell$ is
the length of the chain. Setting $k_BT\sim \hbar \Omega_D\sim \hbar c_s/a$ and
$\ell\sim 1$~mm we get the conclusion that the latter cofactor in
expression~\eqref{eq:15D1} gets the order of $10^7$. In fact, the considered
oscillator chain simulating one-dimensional defects of crystal lattice is
rather formal. Real atoms forming such defects embedded in solid bulk cannot
deviate from their equilibrium positions for distances exceeding the lattice
spacing. So we could expect that exactly sufficiently long low-dimensional
defects limit the accuracy of the dynamical nanostandard on scale about
angstroms on the side of thermal solid vibrations.

The work was supported in part by Grant 04-02-81059 of the joint RFBR-WRFBR collaboration.

\end{document}